\documentclass[journal,a4paper]{IEEEtran}

\usepackage{cite}

\usepackage{bbm}
\usepackage{amsfonts,latexsym,amsmath}
\usepackage{epstopdf}
\usepackage{amssymb}
\usepackage{tikz}
\usetikzlibrary{positioning,shadows,arrows}
\usepackage{caption}
\usepackage{array}

\usepackage{eqparbox}

\usepackage[tight,footnotesize]{subfigure}
\newcommand{\dd}{\,d}

\def\vec#1{{\bf #1}}

\def\blambda{\mbox{\boldmath$\lambda$}}

\def\sgn{\mbox{sgn}}

\def\bH{{\bf H}}
\def\bU{{\bf U}}
\def\bV{{\bf V}}
\def\bI{{\bf I}}

\def\bR{{\bf R}}
\def\bS{{\bf S}}
\def\bT{{\bf T}}
\def\bZ{{\bf Z}}

\def\ba{{\bf a}}
\def\bc{{\bf c}}
\def\be{{\bf e}}

\def\bk{{\bf k}}

\def\bn{{\bf n}}

\def\bs{{\bf s}}

\def\bx{{\bf x}}
\def\by{{\bf y}}

\def\bv{{\bf v}}

\def\bZero{{\mathbf 0}}

\def\blambda{\mbox{\boldmath$\lambda$}}

\def\bsigma{\mbox{\boldmath$\sigma$}}

\def\bDelta{{\mathbf \Delta}}

\newfont{\bb}{msbm10 scaled 1100}

\def\bara{{\bar a}}
\def\ac{{a_c}}
\def\barac{{\bar{a}_c}}
\def\baraz{{\bar a}_z}
\def\barb{{\bar b}}
\def\bc{{b_c}}
\def\barbc{{\bar{b}_c}}
\def\barbz{{\bar b}_z}

\def\ii{\mathrm{i}}

\def\cE{{\mathcal E}}

\usepackage{amsthm}
\theoremstyle{plain}
\newtheorem{theorem}{Theorem}

\newtheorem{corollary}[theorem]{Corollary}

\begin{document}
\bstctlcite{BSTcontrol}
\title{Outage Capacity for the Optical MIMO Channel}

\author{Apostolos~Karadimitrakis$^{1}$, Aris~L.~Moustakas$^{1,2}$ and Pierpaolo Vivo$^3$
\thanks{(1): Department of Physics, Universtity of Athens, Greece}
\thanks{(2): Supelec, Gif sur Yvette Cedex, France}
\thanks{(3): Laboratoire de Physique Th\'{e}orique et Mod\'{e}les Statistiques, UMR CNRS 8626, Universit\'{e} Paris-Sud, 91405 Orsay, France.}
\thanks{This paper was submitted in part at the IEEE ISIT in July 2013 under the title ``Large Deviation Approach to the Outage Optical
MIMO Capacity''. ALM is the recipient of DIGITEO Chair ``ASAPGONE''. PV acknowledges financial support from project  Labex PALM-RANDMAT.}}

\maketitle


\begin{abstract}
MIMO processing techniques in fiber optical communications have been proposed as a promising approach to meet increasing demand for information throughput. In this context, the multiple channels correspond to the multiple modes and/or multiple cores in the fiber. In this paper we characterize the distribution of the mutual information with Gaussian input in a simple channel model for this system. Assuming significant cross talk between cores, negligible backscattering and near-lossless propagation in the fiber, we  model the transmission channel as a random complex unitary matrix. The loss in the transmission may be parameterized by a number of unutilized channels in the fiber. We analyze the system in a dual fashion. First, we evaluate a closed-form expression for the outage probability, which is handy for small matrices. We also apply the asymptotic approach, in particular the Coulomb gas method from statistical mechanics, to obtain closed-form results for the ergodic mutual information, its variance as well as the outage probability for Gaussian input in the limit of large number of cores/modes. By comparing our analytic results to simulations, we see that, despite the fact that this method is nominally valid for large number of modes, our method is quite accurate even for small to modest number of channels.
\end{abstract}


\IEEEpeerreviewmaketitle

\section{Introduction}

\IEEEPARstart{T}{he} ongoing exponential growth in wire-line data traffic is primarily driven by high-bandwidth digital applications, such as video-on-demand, cloud computing and tele-presence. As a result, it is expected that the currently deployed infrastructure will soon reach its limits, leading to the so-called ``capacity crunch'' \cite{Scaling(Crunch)10}. To counter this trend, scientists have been working towards exhausting all available degrees of freedom of fiber-optical transmission, including the bandwidth (through WDM modulation), available power (subject to power constraints imposed by non-linearities), and polarization diversity \cite{Winzer2011_OpticalMIMOCapacity}. One possibility to increase throughput is spatial modulation, which would allow multiple transmission streams within the same fiber or fiber bundle. This can be achieved by designing multi-mode (MMF) and/or multi-core fibers (MCF).

An important issue that arises is that typically there is cross-talk between fiber modes, which increases with segment length \cite{Takenaga10CrosstalkMCF} and can be attributed to imperfections, as well as to the twist and the bending of the fiber \cite{Fini10CrosstalkBentMCF,Hayashi11LowCrosstalkMCF}, and slight variations in the local temperature \cite{Kato00TempChromDispers}. There have been two trends of work in this direction. In the first, effort has been made to minimize cross talk between cores to extremely low levels \cite{Zhu2010_7coreMCF}, thus not having to deal with self-interference. While appealing from a signal processing point of view, the downside is that this methodology does not scale, in the sense that coupling becomes unavoidable with increasing number of cores in a fiber. Another more pragmatic approach is to design fibers without bothering about the appearance of cross-talk. Indeed, bringing cores close to each other can lead to power being spread at the receiver side evenly in the outlook of the channel \cite{Morioka2012_CommMag_Enhancing_OCommsFibers}.

%
Recently, it was proposed \cite{Winzer2011_OpticalMIMOCapacity, TarighatFundamentals07, Hsu06capacityenhancement} to use sophisticated transceiver techniques developed in  the context of wireless communications between multiple transmitting and receiving antennas (MIMO), which can mitigate self-interference, thus providing significant throughput increases. Of course, optical fiber multi-core systems have several differences compared to multi-antenna wireless systems, which need to be addressed. One important difference is the one-dimensional, near-lossless propagation through the optical fiber. As a result, the incoming and outgoing propagating modes of the fiber are related through a so-called scattering matrix \cite{Jackson_EM_book}, which is unitary in the limit of lossless propagation. In contrast, since wireless propagation incurs significant radiation loss to the environment, the corresponding channel coefficients may be taken to be i.i.d and Gaussian \cite{Simon2006_UnitaryPaper, Forrester2010_book}. Second, due to the existence of non-linearities at high powers, one should specifically have in mind low to moderate total power. Third, in contrast to the wireless setting, where due to physical motion the channel fades significantly over time, the variability of the channel is mostly over different frequencies and fiber segments. Hence, given that a given packet is likely to travel over different segments and frequencies, which cannot be known apriori to the transmitter, it is important to define an outage criterion over the realization of the channel matrix in this context. Finally, the practical metric for the performance is not the ergodic mutual information, but, rather, the outage capacity at very low outage (e.g. $10^{-4}$) \cite{Winzer2011_OpticalMIMOCapacity}, due to the fact that feedback from the receiver to the transmitter to request a retransmission in the case of packet loss, is almost always impossible.

It is therefore important to develop a propagation channel model for the fiber-optical MIMO channel, which addresses these issues. Several attempts in this direction have been made \cite{Shah05coherentoptical, Hsu06capacityenhancement}, however the unitary aspect of propagation has not been dealt with explicitly until \cite{Winzer2011_OpticalMIMOCapacity} and then \cite{Dar2012_JacobiMIMOChannel} introduced the unitary aspects of the transmission channel. In particular, \cite{Dar2012_JacobiMIMOChannel} introduced in a somewhat ad-hoc fashion the so-called Jacobi MIMO channel, in which the matrix corresponding to the channel is a rectangular submatrix from a Haar distributed random matrix from $U(N)$.

In this paper, we derive a channel model for an fiber-optical MIMO channel when the coupling between transmitting channels is strong and backscattering is weak. These two assumptions allow us to provide the general random matrix that characterizes the propagation in such a multimode fiber in the presence of time-reversal symmetry. The resulting model is similar to the one introduced by \cite{Dar2012_JacobiMIMOChannel}, but it also parameterizes loss in the fiber. We show how for increased loss, the channel interpolates between unitary and Gaussian. This channel allows us to analyze the outage capacity of the optical MIMO channel. As mentioned above, this is the relevant information transmission metric for fiber-optical coupled multi-core channels. We analyze the characterize the problem in a dual way. First, we obtain closed form expressions for the outage probability for small numbers of channels. We also obtain analytical expressions, which are valid technically in the limit of large channel numbers, but also work well over smaller channel numbers. It is particularly suited to obtain outage mutual information for very low outages with finite SNR. Essentially, it amounts to calculating the rate function of the logarithm of the average moment generating function of the mutual information. The methodology we use is based on the so-called Coulomb gas approach which was developed in the physics literature in the context of random matrix theory \cite{Dyson1962_DysonGas} in the 60's. It is quite intuitive because it interprets the eigenvalues as point charges on a line repelling each other logarithmically. The Coulomb gas method has seen recently a renewed interest in its use to obtain large deviations results for random matrix problems \cite{Majumdar2006_LesHouches, Vivo2007_LargeDeviationsMaxEigvalueWishart, Dean2008_ExtremeValueStatisticsEigsGaussianRMT, Vivo2008_DistributionsConductanceShotNoise} and also in communications \cite{Kazakopoulos2011_LivingAtTheEdge_LD_MIMO, Chen2012_CoulombMIMO}. We will follow the basic steps discussed in more details in \cite{Kazakopoulos2011_LivingAtTheEdge_LD_MIMO}. As a by-product of this analysis we obtain the ergodic mutual information and its variance for this channel.

\subsection{Outline}

In the next section we will define the system model, and show that that the appropriate channel matrix is a random Haar unitary matrix and also define the outage probability, which we would like to analyze. In Section \ref{sec:ExactSolution} we provide the equations describing the closed-form exact solution for the outage probability, details of which are given in Appendix \ref{app:ExactSolution}. In Section \ref{sec:CoulombGasMethodology} we introduce the mathematical methodology of the Coulomb gas and provide our analytic results. Section \ref{sec:Simulations} deals with numerical validation of our results.
Finally, in Section \ref{sec:Conclusion} we conclude.

\section{System Model}
\label{sec:SystemModel}

We consider a single-segment $N$-channel lossless optical fiber system, with $N_t\leq N$ transmitting channels excited and $N_r\leq N$ receiving channels coherently excited in the input (left) and output (right) side of the fiber. The propagation through the fiber may be analyzed through its $2N\times 2N$ scattering matrix given by \cite{Beenakker1997_MesoscopicReview, Winzer2011_OpticalMIMOCapacity}
\begin{equation}
\mathbf{S}=
\begin{bmatrix}
\mathbf{R}_{\ell}&\mathbf{T}_\ell \\
\mathbf{T}_r &\mathbf{R}_{r}
\end{bmatrix}
\label{channel_matrix}
\end{equation}
This matrix ``connects'' the $N$ left ($\ell$) with the $N$ right ($r$) modes of the fiber. The $k$th column (for $k=1,\ldots,N$) of $\mathbf{R}_\ell$ correspond to the reflection coefficients at the left of the $N$ modes of the fiber when a unit amplitude signal is inserted from the $k$th left input of the fiber. The same input signal results to transmission through the fiber, with transmission coefficients at the right hand of the fiber given by the $k$th column of $\mathbf{T}_r$. In an analogous fashion the $k$th columns of $\mathbf{R}_r$ and $\mathbf{T}_\ell$ correspond to the right-reflection and left-transmission coefficients when a unit amplitude signal is inserted from the $k$th right input of the fiber. The input signal is represented by an $2N$ dimensional vector, in which the first $N$ entries correspond to the amplitudes of the left-incoming signal and the remaining entries to the amplitudes of the right-incoming signal.

We now assume that the signal propagates through the above $N$ channels. In this case, for any input $\bv_{in}$ the total input power into the fiber is equal to the total output power, i.e.
\begin{equation}\label{eq:energy_conservation}
  \bv_{in}^\dagger \bv_{in} = \bv_{out}^\dagger \bv_{out} = \bv_{in}^\dagger \bS^\dagger \bS \bv_{in}
\end{equation}
since $\bv_{out}=\bS\bv_{in}$. As a result, the matrix $\bS$ has to be unitary, i.e. $\bS^\dagger\bS = \bI_{2N}$.

A second important property of the scattering matrix relates to its time-reversal symmetry. It is well known that electromagnetic propagation in the absence of external magnetic fields is symmetric under time reversal. In this context, time reversal corresponds to a change in the direction of propagation and time. For example, under time-reversal the amplitude of a  propagating plane-wave $\psi(\bx,t)=\exp[\ii(\bk\bx-\omega t)]$ changes both time $t\rightarrow-t$ and propagation direction $\bk\rightarrow -\bk$. Hence, time-reversal amounts to phase conjugation \cite{Jackson_EM_book}. Therefore, if propagation through the optical fiber is to be time-reversal invariant, feeding the system with the time-reversed version of the output should produce the original version of the input. This implies that $\bR_\ell=\bR_\ell^T$, $\bR_r=\bR_r^T$ and $\bT_\ell=\bT_r^T$. As a result, we are left with three different matrices, namely $\bR_\ell$, $\bR_r$ and $\bT_\ell=\bT_r^T\equiv \bT$. These matrices are not independent, since they share the same singular values, since $\bR^\dagger_\ell\bR_\ell+\bT^\dagger\bT = \bR^\dagger_r\bR_r+\bT^\dagger\bT=\bI_N$. It is convenient to define the matrix $\bDelta$ as the diagonal matrix with the eigenvalues of $\bT^\dagger \bT$. It has been shown elsewhere \cite{Martin1992_Landauer, Beenakker1997_MesoscopicReview} that $\bS$ can be expressed in terms of $\bDelta$ by means of a so-called polar decomposition as follows
\begin{equation}
\bS =
\begin{bmatrix}
\bU&\bZero \\
\bZero &\bV
\end{bmatrix}
\begin{bmatrix}
-\bDelta^{1/2}&(\bI_N-\bDelta)^{1/2} \\
(\bI_N-\bDelta)^{1/2} &\bDelta^{1/2}
\end{bmatrix}
\begin{bmatrix}
\bU^T&\bZero \\
\bZero &\bV^T
\end{bmatrix}
\label{eq:polar_decomposition}
\end{equation}
As a result, the information of the scattering matrix $\bS$ is encoded in the matrices $\bDelta$, $\bU$ and $\bV$.

We now discuss two important properties of the scattering matrix as seen from experimental data in the literature, which will help describe it better. We start with the strength of backscattering, i.e. reflection in optical fibers. This process is typically due to localized imperfections in the fiber and is sometimes called Rayleigh scattering. The strength of the reflected light is typically proportional to the product of the density of such imperfections and the length of the fiber \cite{Gysel1990_StatisticalPropertiesRayleighScettering}, i.e. proportional to the average number of such imperfections over the fiber length travelled. Due to the high quality of fiber production techniques this imperfection density is extremely small. Hence, in \cite{Gifford2012_OpticalBackscatterReflectometry} single core fibers have reflection coefficients approximately equal to -120dB/mm, which amounts to -30dB per 1000km. Similarly, in \cite{Martinez2010_BackscateredOpticalPowerBidirectionPON} a 25km single mode fiber has Rayleigh backscattered power roughly -30dB. These very low reflected powers appear in single mode fibers, however, we conjecture that they should be quite low for multi-core fibers described below. As a result of this low backscattering amplitudes we may assume that the reflection in the fiber may discarded, and hence $\bDelta\approx\bZero$.

A second important property of the scattering matrix in a multicore/multimode fiber is the considerable mixing between core transmissions. For example, in a 60km three coupled core fiber analyzed in \cite{Morioka2012_CommMag_Enhancing_OCommsFibers}, the crosstalk is so strong that light injected into one core is equally distributed across all cores in the output. Considerable crosstalk has been seen in other cases, e.g. in \cite{Ohasi2012_BackscatteredPower_CrosstalkMultiCoreFibers} where crosstalk of -25dB/km was observed. Even if this effect is smaller that in \cite{Morioka2012_CommMag_Enhancing_OCommsFibers} above due to the different design of the cores (it results to -8dB coupling for 60km), it highlights the relevance and ubiquitous nature of crosstalk in multicore fibers, when they have their cores placed close to each other. It should be pointed out that the difference in magnitude of backscattering and crosstalk can be attributed to different mechanisms being responsible for the two effects. In the backscattering case, as discussed above, the effect is due to localized scattering\cite{Gifford2012_OpticalBackscatterReflectometry}, while in the latter the mechanism is scattering among the core modes due their proximity, or due to bending \cite{Fini10CrosstalkBentMCF}.

\subsection{Statement of Problem}
\label{sec:ProblemStatement}

In summary, we consider fibers with negligible backscattering and strong mixing between core modes. We assume this mixing to be random over different frequency subbands, due to strong delay spread. For example, in \cite{Ryf2013_CombinedSDMWDM_FMF} 10nsec delay spreads were measured over 700km transmission over a 6 mode fiber using 50GHz sub-band widths. Hence the transmission matrix $\bT$ will be modelled as a Haar random matrix of dimension $N\times N$. Without loss of generality we assume $N_t\leq N$ transmitting channels and $N_r\leq N$ receiving channels, and therefore we only consider a submatrix of the full transmission matrix, which we denote by $\bU$, since not all transmitting or receiving channels may be available to a given link. For simplicity we assume that this is the upper left corner of $\bT$. We should emphasize that the remaining $N-\max(N_t,N_r)$ ``untapped'' channels in $\bT$ can be used to model {\em loss} in the fiber propagation \cite{Simon2006_UnitaryPaper}. Indeed, in the limit of large $N\gg N_t,N_r$ the channel will converge to a Gaussian distributed channel,\cite{Simon2006_UnitaryPaper} similar to the case of open space wireless propagation, where the signal loss is significant. As a result, the corresponding MIMO channel for this system reads
\begin{equation}
 {\bf y} = \mathbf{U} {\bf x} + {\bf z}
 \label{system}
\end{equation}
with coherent detection and channel state information only at the receiver \cite{Foschini1998_BLAST1, Telatar1995_BLAST1}. ${\bf x}$, ${\bf y}$ and ${\bf z}$ are the $N_t\times 1$ input, the $N_r\times 1$ output signal vectors and the $N_r\times 1$ unit variance noise vector, respectively, all assumed for simplicity to be complex Gaussian. This assumption is also based on the optical MIMO modulation scheme, which uses MZM (\textit{Mach-Zehnder Modulator}) to modulate a continuous wave (CW) laser to generate the digital signal, which is then, transmitted through the fiber. This modulation is achieved by equally splitting the incoming optical signal and enforcing a time delay (phase shift) in one path, before recombining it. We also assume no mode-dependent loss. As a result, the mutual information can be expressed as
\begin{eqnarray}
I_{N}({\bf U})&=&\frac{1}{N_t}\log \det(I + \rho {\mathbf{U}^{\dagger} \mathbf{U}}) \\ \nonumber
&=& \frac{1}{N_t}\sum_{k=1}^{N_t} \log(1+\rho\lambda_k) \\ \nonumber
&=& \int_0^1 p(x)dx \log(1+\rho x)
\end{eqnarray}
In the last equation $p(x)$ is the spectral density of $\bU^\dagger \bU$ defined as
\begin{equation}\label{eq:spec_density_def}
p(x)=\frac{1}{N_t}\sum_k \delta(x-\lambda_k)
\end{equation}
Also, ``$\log$'' is the natural logarithm, $ \rho $ is the average total signal-to-noise ratio, $\lambda_k$ are the eigenvalues of the matrix ${\bf U}^\dagger {\bf U}$ and we assume for concreteness $N_t\leq N_r$. It is useful to define $\beta=N_r/N_t>1$, $N_0=N-N_t-N_r$ and $n_0 = N_0/N_t\geq 0$. If $N_0<0$, \cite{Dar2012_JacobiMIMOChannel} showed that we may recover the above form by replacing $N_t\rightarrow N-N_r$, $N_r\rightarrow N-N_t$ and $N_0\rightarrow -N_0$ and $I_N\rightarrow I_N+n_0\log(1+\rho)$. It should be emphasized that the above mutual information is used as a performance metric of the channel.

We may now define the main problem we address, namely the calculation of
\begin{eqnarray}\label{eq:Pout_def}
  P_{out}(r) &=& Prob(I_N< r) \\
   &=& E_{{\bf U} } \left[\Theta(r-I_N({\bf U}))\right]
\end{eqnarray}
where $\Theta (x)$ is the indicator (step) function. We will also analyze the density of $r$ i.e.
\begin{equation}\label{eq:P(r)}
  P(r)=P_{out}'(r)=E_{\mathbf{U}}[\delta(r-I_N(\mathbf{U}))]
\end{equation}
The aim of this paper is to calculate the tails of the distribution of the rate $r$.
The first step is to express the joint distribution of eigenvalues of ${\bf U}^\dagger {\bf U}$ as derived initially in \cite{Simon2006_UnitaryPaper} and more recently in this context \cite{Dar2012_JacobiMIMOChannel}
\begin{eqnarray}
P_{\blambda}(\lambda_{1}...\lambda_{N_t}) = {\cal Z}_{N_t}^{-1} \prod\limits_{n<m}| \lambda_{n}-\lambda_{m}| ^2 \prod\limits_{k} \lambda_{k}^{| N_{t}-N_{r}|} (1-\lambda_{k})^{N_{0}}
\label{eq:Plambda_def}
\end{eqnarray}
In the above, ${\cal Z}_{N_t}$ is a normalization constant defined in \eqref{eq:ZN_def}.

In the above equation, we can see that when $N_0$ becomes large, the last term can be approximated roughly as $(1-\lambda)^{N_0}\approx e^{-N_0\lambda}$. This corresponds to the eigenvalue distribution of a Wishart matrix \cite{Wang2002_OutageMutualInfoOfSTMIMOChannels}, which is typically used to model wireless MIMO channel propagation because the latter has significant power loss in the atmosphere. Hence, it can be seen that the parameter $N_0$ can effectively model power loss through the fiber and provide a continuous cross-over between lossless and lossy fibers \cite{Simon2006_UnitaryPaper}.

In the next section, we will show how the above expression can be used to provide a closed form solution for the outage probability, in terms of finite sums of simple functions. However, it will become clear that for increasing channel numbers, the formula becomes quite cumbersome, without providing much intuition. Hence, in Section \ref{sec:CoulombGasMethodology} a different approach will be adopted, namely the large-$N$ analysis of the outage probability using the Coulomb gas formalism.

\section{Exact Solution}
\label{sec:ExactSolution}

In this section, we will obtain a closed form expression for the outage probability $P_{out}(r)$. We start by introducing an integral representation for the $\Theta$ function
\begin{equation}\label{eq:Theta_def}
1- \Theta(x) = \Theta(-x) = -\int_{-\infty}^{+\infty} \frac{dp}{2\pi \ii} \frac{e^{\ii px}}{p+\ii\epsilon}
\end{equation}
where $\epsilon$ is an infinitesimal positive number indicating that the $k$-integral goes over the pole at zero. As a result, the outage probability can be expressed as follows:
\begin{align}\label{eq:Pout_Fourier_xform1}
&1-P_{out}(r)=\int d\blambda P_{\blambda}(\blambda)  \int_{-\infty}^\infty \frac{dp}{2\pi} \frac{e^{\ii pN_tr}}{\epsilon-\ii p}\prod_{n=1}^{N_t} \left(1+\rho\lambda_n\right)^{-\ii p}
\end{align}
where the integral notation $\int d\blambda$ signifies multiple integration over all $\lambda_k$ for $k=1,\cdots,N_t$. In Appendix \ref{app:ExactSolution} we show how the above multiple integral can be evaluated. The final result can be expressed as follows:
\begin{align}\label{eq:exact_result_final1}
\nonumber  1-P_{out}(r) &=\sum_{\bk,\bn} c_{\bk,\bn}
\sum_{\bsigma}(-1)^{|\bsigma|} \times \\
&\times \sum_{\ell=\ell(r)}^{N_t} (-1)^{\ell+N_t} d_\ell(\bs_{\bsigma}) F(N_t r-\ell\log(1+\rho),\bs_{\sigma})
\end{align}
where the sum of $\bk$ is over $[0,|N_t-N_r|]^{N_t}$,  the sum of $\bn$ is over $[0,N_0]^{N_t}$ and the sum over $\bsigma$ is over all permutations of $(1,\cdots,N_t)$ with signature $|\bsigma|$. The $N_t$-dimensional integer vector $\bs_{\bsigma}$ has components $s_j=j+\sigma_j -1 +k_j+N_0-n_j$ and $\ell(r)$ is the smallest integer for which $N_t r<\ell\log(1+\rho)$, while
\begin{align}
&c_{\bk,\bn} = \frac{N_t!\prod_{j=1}^{N_t} c_{k_j,n_j}}{{\cal Z}_N\rho^{N_t^2+(|N_t-N_r|+N_0)N_t}} \\  %
&d_\ell(\bs) = \be_\ell\left((1+\rho)^{s_1},\cdots,(1+\rho)^{s_{N_t}}\right) \\
&F(z,\bs)= \prod_{j=1}^{N_t} s_j^{-1} + \sum_{j=1}^{N_t}\frac{e^{s_j z}}{s_j\prod_{k\neq j}(s_k-s_j)}
\end{align}
In the above $c_{k,n}$ are given in \eqref{eq:c_kn}, ${\cal Z}_{N_t}$ is given in \eqref{eq:ZN_def}, while $\be_\ell(x_1,x_2,\cdots,x_{N_t})$ is the elementary symmetric polynomial of degree $\ell$\cite{Macdonald1995_SymmetricFunctions_book}. The prescription of how to deal with $F(z,\bs)$ in the case where two or more integers $s_i$ are equal is discussed in Appendix \ref{app:ExactSolution}. We also note that the density of $r$, $P(r)$, can be obtained directly from the above by differentiation with respect to $r$.

Although analytic and in closed form, the above result is handy and provides intuition for the answer at best for small values of $N_t$, $N_r$, $N_0$. When this is not the case, one needs an alternate path, which can be achieved using the asymptotic approach in $N$, which will be discussed next.

\section{Coulomb Gas Methodology}
\label{sec:CoulombGasMethodology}

In this section we will follow a complementary approach to the above and will derive the outage probability in the limit of large channel numbers. The first step is to rewrite the joint distribution of eigenvalues of ${\bf U}^\dagger {\bf U}$ provided in \eqref{eq:Plambda_def} in the following form
\begin{eqnarray}
P_{\blambda}(\lambda_{1}...\lambda_{N_t}) &=& {\cal Z}_{N_t}^{-1}e^{-N_t^{2}E(\blambda)}
\label{eq:Plambda_def2}
\end{eqnarray}
where
\begin{IEEEeqnarray}{rCl}\label{eq:energy_fun}
E(\blambda)&=&-\frac{N_{0}}{N_t^2}\sum_{k=1}^{N_t} \log(1-\lambda_k) -\frac{N_r -N_t}{N_t^2}\sum_{k=1}^{N_t} \log(\lambda_k)\nonumber \\
&-&\frac{1}{N_t^2}\sum_{k\neq k'}\log{|\lambda_k-\lambda_{k'}|}
\end{IEEEeqnarray}
$E(\blambda)$ represents the normalized potential energy of $N_t$ unit charges bound on the unit interval $x\in(0,1)$, while repelling from each other and from the boundaries logarithmically. It is reasonable to expect that when $N$ is large, the charges will coalesce to a smooth density $p(x)$. This hypothesis, which is originally due to Dyson \cite{Dyson1962_DysonGas}, and is called the {\em Coulomb (or Dyson) gas approach}, has been used extensively in statistical physics \cite{Mehta_book, Majumdar2006_LesHouches, Forrester2010_book} and more recently in communications \cite{Kazakopoulos2011_LivingAtTheEdge_LD_MIMO}. This hypothesis was set in a more mathematical footing by \cite{BenArous1997_LDWignerLaw} who proved that the large deviations of the law of the spectral density $p(x)$ can be described by a rate function corresponding to the continuum limit of $E(\blambda)$. \cite{BenArous1997_LDWignerLaw} showed this for the case of the Wigner Gaussian matrices, while \cite{HiaiPetz1998_LargeDeviationsWishartEigenvalues} generalized it to Wishart matrices. Their proof can be directly applied to this model by restricting the support of eigenvalues from $\lambda\in(0,\infty)$ to the unit interval $\lambda\in(0,1)$. We will apply this formalism to obtain the tails of $P_{out}(r)$. The first result is summarized in the following theorem, which is proved in Appendix \ref{app:proof_LD}. Let us first denote by ${\cal X}$ the space of probability measures on $(0,1)$, endowed with weak topology.
\begin{theorem}[Large Deviations of Eigenvalue Density]
\label{thm:Coulomb_gas}

Let
\begin{IEEEeqnarray}{rCl}\label{eq:E(p)}
\mathcal{E}[p]&=&-n_{0}\int p(x)\log(1-x)d x -(\beta -1)\int p(x)\log(x)d x \nonumber \\
&-&\iint\limits p(x)p(y)\log{|x-y|}dy dx
\end{IEEEeqnarray}
where $p(x)\in{\cal X}$. Then
\begin{enumerate}
\item ${\cal E}[p]$ is convex on ${\cal X}$
\item ${\cal E}[p]$ obtains its minimum value denoted by ${\cal E}_0$ at a unique probability density $p_0(x)$ on $(0,1)$.
\item $\lim_{N_t\rightarrow \infty} \frac{1}{N_t^2}\log P(I_N\leq r)={\cal E}_0-\inf_{p\in{\cal X}_r} {\cal E}[p]$ where
\begin{eqnarray}\label{eq:Xr_def}
\mathcal{X}_r = \left\{p\in {\cal X} \text{ and } \int_0^1\! p(x)\log(1+\rho x) \dd x \leq \ r\right\}
\end{eqnarray}
\end{enumerate}
\end{theorem}
In this paper we mostly interested in the outage probability defined in \eqref{eq:Pout_def} and therefore the above result is of interest. However, an analogous result can be obtained for the $1-P_{out}(r)$:
\begin{corollary}\label{cor:Coulomb_gas_complement}
If $\mathcal{X}_r$ includes the density $p_0(x)$ (or is arbitrarily close to it), then from the above we conclude that $\inf_{p\in{\cal X}_r} {\cal E}[p]={\cal E}_0$ and hence $\log P(I_N\leq r)/N_t^2\rightarrow 0$. Hence, in this case, we do not strictly speaking have a large deviation event. Nevertheless, in this case it can be shown that the complement of ${\cal X}_r$, namely
\begin{equation}\label{eq:Xr_complement_def}
\mathcal{X}^c_r = \left\{p\in {\cal X} \text{ and } \int_0^1\! p(x)\log(1+\rho x) \dd x > \ r\right\}
\end{equation}
is a large deviation event, i.e.
\begin{equation}\label{eq:LD_complement}
\lim_{N_t\rightarrow \infty} \frac{1}{N_t^2}\log P(I_N > r)={\cal E}_0-\inf_{p\in{\cal X}^c_r} {\cal E}[p]
\end{equation}
\end{corollary}

Due to the convexity of ${\cal E}[p]$ and ${\cal X}_r$, it is sufficient to find a local minimum of the functional, subject to the constraints, which then is ensured to be a global minimum. One handy way to do so is to introduce a Lagrangian and include the constraints of normalization and positivity of $p(x)$ using Lagrange multipliers. We thus have
\begin{eqnarray}
\label{eq:L0}
{\mathcal L}_{0}[p,\nu,c]	&=& {\mathcal E}[p] - c\left(\int_0^1\!\!\! p(x) \dd x -1\right) \nonumber\\
					&-& \int_0^1 \!\!\!\nu(x)p(x) \dd x
\end{eqnarray}
\begin{eqnarray}
\label{eq:L1}
{\mathcal L}[p,\nu,c,k]	&=& {\mathcal L}_0[p,\nu,c]\nonumber\\
					&-& k\left(\int_0^1\!\!\! p(x)\log(1+\rho x) \dd x-r\right)
\end{eqnarray}
from which we obtain ${\mathcal E}_{0}$ and ${\mathcal E}(r)$ by maximizing over the dual parameters $\nu$ (non-negativity constraint), $c$ (normalization constraint) and $k$ (mutual information constraint):
\begin{eqnarray}
\label{eq:minimum_0}
{\mathcal E}_0		&=& \sup_{\nu\geq 0; \, c} \inf_p {\mathcal L}_0[p,\nu,c]\\
{\mathcal E}(r)	&=& \sup_{\nu\geq 0; \, c,k} \inf_p {\mathcal L}[p,\nu,c,k]
\label{eq:minimum_1}
\end{eqnarray}
As a result, for large $N_t$ we have
\begin{equation}\label{eq:P_N(r)}
Prob(I_N\leq r) \sim e^{-N_t^{2} \left(\mathcal{E}(r) - \mathcal{E}_{0}\right)}
\end{equation}
The convexity of ${\mathcal L}_0$, ${\mathcal L}$ over $p$ ensures uniqueness of the minimizing $p$.  Therefore, if we find a local minimum for the corresponding Lagrangian for $k$, $c$ and $\nu$ that satisfy the constraints, this will be a unique one.

It is also worth pointing out that the only difference between ${\mathcal E}_0$ and ${\mathcal E}(r)$ above is that the former can be seen as the maximum over ${\mathcal L}[p,\nu,c,k]$ keeping $k=0$; this relation will come in handy later, because it allows us to work with ${\mathcal L}$ and at the very last step set $k=0$ to obtain ${\mathcal E}_0$. This result is in agreement with (\ref{eq:appE0}) derived in Appendix \ref{app:E_0} using other methods.  To find a local minimum of ${\mathcal L}$, it suffices to calculate its functional derivative with respect to $p$ and which is then set to zero. Note that the functional derivative of ${\mathcal L}$ at $p\in{\cal X}_{r}$ is the distribution $\delta{\mathcal L}[p, \nu, c, k]$ whose action on test functions which leave ${\cal E}[p]$ finite is given by:
\begin{equation}
\label{eq:funcderiv}
\left\langle\delta{\mathcal L}[p],\phi\right\rangle
			= \frac{d}{dt}\bigg|_{t=0} \!\!{\mathcal L}[p + t\phi].
\end{equation}
Note that maximizing the result with respect to $k$ and $c$ simply corresponds to enforcing the normalization and mutual information constraints that appear in (\ref{eq:L0}) and (\ref{eq:L1}):
\begin{eqnarray}
\label{eq:norm_condition}
\int_0^1 p(x)\dd x &=& 1 \\
\int_0^1 p(x) \log(1+\rho x)\dd x &\leq& r
\label{eq:mutual_information_condition}
\end{eqnarray}
It is worth commenting here that since we will see that ${\cal E}(r)$ is an decreasing function of $r$ for $r<r_{erg}$ the mutual information constraint becomes essentially an equality constraint, since the infimum of ${\cal E}[p]$ with respect to densities of different mutual information values is obtained at the maximum possible value allowed. The opposite holds for the case $r>r_{erg}$, when ${\cal E}(r)$ is increasing function of $r$. In this case the infimum is over the set ${\cal X}_r^c$ so once again the optimal value is $r$.

The maximization over $\nu(x)$ ensures the non-negativity of $p(x)$. It can be shown \cite{Boyd_book} that either $\nu(x)$ or $p(x)$ are non-zero, therefore making $\nu(x)p(x)=0$. For simplicity we will not analyze this constraint, instead enforcing it explicitly. Setting the functional derivative of ${\cal L}[p]$ to zero results to
\begin{eqnarray}
  2\int_a^b p(x')\log|x-x'| dx' &=& -k\log(1+\rho x) -c \\ \nonumber
  &-&n_0\log(1-x)-(\beta-1)\log(x)
\label{eq:funct_deriv=0}
\end{eqnarray}
for all $x$ in the support of $p(x)$, which is assumed for the moment to be the (connected) interval $(a,b)\subseteq (0,1)$, enforced by $\nu(x)$. Taking the derivative with respect to $x$ in the above we obtain the following integral equation, which has the physically intuitive meaning of force balancing at the charges in $x$:
\begin{equation}\label{eq:Tricomi_eq}
  2{\cal P}\int_a^b \frac{p(x')}{x-x'}dx' = \frac{n_0}{1-x}-\frac{\beta-1}{x} -\frac{k\rho}{1+\rho x}
\end{equation}
where ${\cal P}$ represents the Cauchy principal value of the integral. Once $p(x)$ has been determined, we can obtain $\cE(r)$ by direct integration. To evaluate the double integral in (\ref{eq:E(p)}) we can one integral in terms of (\ref{eq:funct_deriv=0}). Then the value of $c$ can be determined by calculating (\ref{eq:funct_deriv=0}) at $x'=a$ \cite{Vivo2007_LargeDeviationsMaxEigvalueWishart, Kazakopoulos2011_LivingAtTheEdge_LD_MIMO}. Following Tricomi's theorem \cite{Tricomi_book_IntegralEquations, Kazakopoulos2011_LivingAtTheEdge_LD_MIMO} this integral equation may be solved to yield the following general expression
\begin{eqnarray}
\label{eq:gen_solution_int_eq0}
    p(x) &=& \frac{\frac{n\sqrt{(1-a)(1-b)}}{1-x} - \frac{k\sqrt{(1+a\rho)(1+b\rho)}}{1+\rho x} -\frac{(\beta-1)\sqrt{ab}}{x} + C}{2\pi\sqrt{(x-a)(b-x)}}
\end{eqnarray}
where $C$ is a constant. This is a valid solution if the right hand side expression of (\ref{eq:Tricomi_eq}) is $L^{1+\epsilon}$ integrable (for some $\epsilon>0$) over the support $(a,b)$. Clearly, this is not the case if $a=0$ or $b=1$, whenever $\beta>1$ or $n>0$, respectively. Therefore, in those cases the values of $a$ and $b$ need to be found self-consistently, by demanding that $p(x)$ is continuous at that value, i.e. that $p(a>0)=0$ or $p(b<1)=0$. As a result, we find four types of solutions, depending on whether $a=0$ and/or $b=1$. Before summarizing the solution results for these four cases, we obtain the solution for the case $k=0$, which corresponds to most probable value of $r=r_{erg}$. In this case, the eigenvalue distribution that minimizes $\mathcal{L}_0$ is simply
\begin{equation}\label{eq:MP_0}
  p_0(x) = \frac{\sqrt{(x-a_0)(b_0-x)}}{2\pi x(1-x)}
\end{equation}
where
\begin{equation}\label{eq:a0_b0_def}
a_0,b_0=\frac{\left(\sqrt{1+n_0} \pm \sqrt{\beta(n_0+\beta)}\right)^2}{n_0+1+\beta}
\end{equation}
which has been obtained using other methods in \cite{Simon2006_UnitaryPaper, Debbah2003_UnitaryAsymptoticallyFreeMatrices}.
From the above $p_0(x)$, $\cE_0$ can be evaluated. The result thus obtained matches the result obtained using a more direct method in Appendix \ref{app:E_0}.

In the next sections we will obtain the solution for $\Delta {\mathcal E}(r)$ for all allowed values of parameters $n$, $\beta$, $r$. The analysis is based in the methodology in \cite{Kazakopoulos2011_LivingAtTheEdge_LD_MIMO}. It should be stressed that given the convexity of ${\cal E}[p]$ with respect to $p$, it is sufficient to find an acceptable solution of the constrained extremization procedure discussed above. Below we will analyze the four possible types of solutions, corresponding to $a=0$ or $a>0$ and $b=1$ or $b<1$. We will see that for any parameter value of $n$, $\beta$, $r$, there is a single solution to the Tricomi equation above \eqref{eq:Tricomi_eq}, which is consistent with all constraints, as well as positivity and continuity on $(0,1)$. We will see that while continuity will exclude some types of solutions, e.g. $a=0$ when $\beta>1$ and $b=1$ when $n>0$, we will find two or three types of solutions applicable for a given set of $n$ and $\beta$. Of course, only one is valid for any given value of $r$. We will see that there is a critical value of $r$, at which one type of solution becomes invalid, while another becomes applicable. This phase transition is characterized with the attachment of the support of $p(x)$ to a boundary of $(0,1)$ and has been in the literature with a third order phase transition and the Tracy-Widom law \cite{Majumdar2006_LesHouches, Dean2008_ExtremeValueStatisticsEigsGaussianRMT, Vivo2007_LargeDeviationsMaxEigvalueWishart, Kazakopoulos2011_LivingAtTheEdge_LD_MIMO}. In Table \ref{table:solution_types} we summarize the validity of each solution type, denoted by $S_{01}$, $S_{a1}$, $S_{0b}$ and $S_{ab}$, where the first index describes the infimum of the support ($0$ if $a=0$ and $a$ if $a<1$) and the second corresponds to its supremum ($1$ if $b=1$ and $b$ if $b<1$).

\begin{table}[h]
\centering
\begin{tabular}{|c|c|c|c|c|}
  \hline
                   &$S_{0b}$ & $S_{ab}$ &$S_{01}$ & $S_{a1}$\\
  \hline
                   & $a=0$   & $a>0$    & $a=0$   & $a>0$ \\
                   & $b<1$   & $b<1$    & $b=1$   & $b=1$ \\
                   \hline
  $n=0$; $\beta=1$ & $r<r_{c1}$ & -- & $r_{c1}<r<r_{c2}$ & $r>r_{c2}$ \\
  $n>0$; $\beta=1$ & $r<r_{c3}$ & $r>r_{c3}$ & -- & -- \\
  $n=0$; $\beta>1$ & -- & $r<r_{c4}$ & -- & $r>r_{c4}$ \\
  $n>0$; $\beta>1$ & -- & all r & -- & -- \\
  \hline
\end{tabular}
\caption{Summary of validity of four types of solutions depending on the values of $n$, $\beta$ and $r$.}
\label{table:solution_types}
\end{table}

\subsection{Solution $S_{01}$: $a=0$, $b=1$}
\label{sec:S_01}

We start with the most trivial type of solution, namely when the support boundaries $a=0$ and $b=1$ are enforced. This solution can be valid only when $n=0$ and $\beta=1$, since otherwise the right-hand-side of (\ref{eq:Tricomi_eq}) and hence also $p(x)$ \cite{Tricomi_book_IntegralEquations} will not be $L^{1+\epsilon}$-integrable. The resulting optimal normalized spectral density is
\begin{eqnarray}
p(x)=\frac{(z+x)(k+2)-k\sqrt{z(z+1)}}{2\pi(z+x)\sqrt{x(1-x)}}
\end{eqnarray}
The resulting relation between $r$ and $k$ obtained by enforcing the rate constraint is (\ref{eq:mutual_information_condition})
\begin{equation}
\label{eq:rate_constraint_n=0beta=1}
r= r(k)\equiv \log\frac{(1+\sqrt{1+\rho})^2}{4} + k\log\frac{(1+\sqrt{1+\rho})^2}{4\sqrt{1+\rho}}
\end{equation}
and the corresponding value of the exponent $\Delta\mathcal{E}={\cal E}(r)-{\cal E}_0$ becomes quadratic
\begin{IEEEeqnarray}{rCl}
\Delta\mathcal{E}=\frac{\left(r-2\log\frac{1+\sqrt{1+\rho}}{2}\right)^2}{2\log\frac{(\sqrt{z}+\sqrt{z+1})^2}{4\sqrt{z(z+1)}}}
\end{IEEEeqnarray}

The validity of the above result breaks down when the positivity constraint of $p(x)$ is violated. This happens when $k<k_{c1}$ or $k>k_{c2}$, where
\begin{equation}
k_{c1} = -\frac{2\sqrt{z+1}}{\sqrt{z+1}-\sqrt{z}},~~~~~~k_{c2}=\frac{2\sqrt{z}}{\sqrt{z+1}-\sqrt{z}}
\label{eq:kc1_kc2_a=0b=1}
\end{equation}
with corresponding values of the rate obtained through $r<r_{c1}=r(k_{c1})$ and $r>r_{c2}=r(k_{c2})$, respectively. If this is true, we need seek for a solution allowing $b<1$, or $a>0$, respectively. This will be analyzed in the next two subsections.

\subsection{Solution $S_{0b}$: $a=0$, $b<1$}
\label{sec:S_0b}

This solution can only be valid for $\beta=1$. In this case the resulting optimal eigenvalue density is given by
\begin{IEEEeqnarray}{rCl}\label{eq:p(x)_a=0,b<1}
p(x) = \frac{1}{2\pi}\sqrt{\frac{b-x}{x}}\left(\frac{n_0}{\sqrt{1-b}}\frac{1}{1-x}-\frac{k\sqrt{z}}{\sqrt{z+b}}\frac{1}{z+x}\right)\quad
\end{IEEEeqnarray}
The normalization condition (\ref{eq:norm_condition}) gives
\begin{equation}
\label{eq:norm_condition_a0b<1}
\frac{n_0}{\sqrt{1-b}}+k\sqrt{\frac{z}{z+b}}=2+n_0+k
\end{equation}
which is shown in Appendix \ref{app:uniqueness_norm_condition} to have a unique solution, while the rate equality (\ref{eq:mutual_information_condition}) condition gives
\begin{IEEEeqnarray}{lCl}
r &=& r(k)\equiv \log(\rho b) + \frac{nb}{2\sqrt{1-b}} \left(G(z/b,0)-G(z/b,-1/b)\right) \nonumber \\
&-& \frac{kb}{2\sqrt{(z+b)z}} \left(G(z/b,0)-G(z/b,z/b)\right)
\end{IEEEeqnarray}
and finally
\begin{IEEEeqnarray}{lCl}
\mathcal{E}(r) &=&\frac{k}{2}\bigg[r-\log{\left(1+b\rho\right)}\bigg]-\frac{n\log(1-b)}{2}-\frac{(n+2)\log{b}}{2} \nonumber \\
&-&\frac{n^2b}{4\sqrt{1-b}}\left(I_3(\frac{1}{b}-1)+G(\frac{1}{b}-1, \frac{1}{b}-1)\right) \nonumber\\
&+&\frac{nkb}{4\sqrt{z(z+b)}}\left(I_3(\frac{1}{b}-1)+G(\frac{1}{b}-1, -1-\frac{z}{b})\right) \nonumber\\
&-&\frac{nb}{2\sqrt{1-b}}\left(I_3(0)+G(0, -1-\frac{1}{b})\right) \nonumber\\
&+&\frac{kb}{2\sqrt{z(z+b)}}\left(I_3(0)+G(0, -1-\frac{z}{b})\right)
\end{IEEEeqnarray}

When $n=0$, (\ref{eq:norm_condition_a0b<1}) breaks down (and hence $p(x)$ is not properly normalized) if $k>k_{c1}$, assuming of course $b\leq 1$. Hence, in this case this solution is invalid in agreement with the discussion in the previous subsection.

In contrast when $n>0$, the above solution breaks down when $p(x)<0$ for small $x$. This happens when, in addition to (\ref{eq:norm_condition_a0b<1}) $n(1+z)<(2+n+k_{c3})\sqrt{1-b_{c3}}$, which corresponds to $r> r_{c3}=r(k_{c3})$. In this case, we need to allow $a>0$, which will be analyzed in a later subsection.

\subsection{Solution $S_{a1}$: $a>0$, $b=1$}
\label{sec:S_a1}

In the spirit of previous subsections, this solution can only be valid when $n=0$. In this case the resulting optimal eigenvalue density is given by
\begin{equation}
p(x)=\frac{\sqrt{x-a}}{2\pi\sqrt{1-x}}\left[k\sqrt{\frac{z+1}{z+a}}\frac{1}{z+x}+\frac{\beta -1}{x}\sqrt{\frac{1}{a}}\right]
\end{equation}
Using the normalization equation
\begin{equation}\label{eq:norm_condition_a>0b=1}
  \beta+1+k=\frac{\beta-1}{\sqrt{a}}+\frac{k\sqrt{z+1}}{\sqrt{z+a}}
\end{equation}
and the rate constraint (\ref{eq:mutual_information_condition})
\begin{IEEEeqnarray}{lCl}
r &=& r(k)\equiv \log(\rho(1-a)) \nonumber \\
&+& \frac{k(1-a)}{2\sqrt{(z+1)(z+a)}} \left(I_3(\frac{a+z}{1-a})+G(\frac{a+z}{1-a},\frac{a+z}{1-a})\right) \nonumber\\
&+&\frac{(\beta-1)(1-a)}{2\sqrt{a}} \left(I_3(\frac{a+z}{1-a})+G(\frac{a+z}{1-a},\frac{a}{1-a})\right)
\end{IEEEeqnarray}
we can finally calculate $\mathcal{E}(r)$.
\begin{IEEEeqnarray}{lCl}
\mathcal{E}(r)&=&\frac{k}{2}\left(r-\log{\left(1+a\rho\right)}\right)-\frac{\beta -1}{2}\log{a} - \frac{\beta +1}{2}\log(1-a) \nonumber \\
&-&\frac{(\beta-1)^2(1-a)}{4\sqrt{a}} \left(I_3(\frac{a}{1-a})+G(\frac{a}{1-a},\frac{a}{1-a})\right) \nonumber \\
&-&\frac{(\beta-1)k(1-a)}{4\sqrt{(z+a)(z+1)}} \left(I_3(\frac{a}{1-a})+G(\frac{a}{1-a},\frac{a+z}{1-a})\right)  \nonumber \\
&-&\frac{(\beta-1)(1-a)}{2\sqrt{a}} \left(I_3(0)+G(0,\frac{a}{1-a})\right)  \nonumber \\
&-&\frac{k(1-a)}{4\sqrt{(z+a)(z+1)}} \left(I_3(0)+G(0,\frac{a+z}{1-a})\right)
\end{IEEEeqnarray}

When $\beta=1$, for $k<k_{c2}$, where $k_{c2}$ is defined in (\ref{eq:kc1_kc2_a=0b=1}), (\ref{eq:norm_condition_a>0b=1}) gives $a<0$, which is obviously not allowed, hence invalidating this solution. This is in agreement with subsection \ref{sec:S_01}.

In contrast when $\beta>1$, the above solution breaks down when $p(x)<0$ for $x\approx1$. This happens when, in addition to (\ref{eq:norm_condition_a>0b=1}) we have $(\beta-1)z+(1+\beta+k_{c4})\sqrt{a_{c4}}=0$, which corresponds to $r= r_{c4}=r(k_{c4})$. In this case, we need to also allow $b<1$, which will be analyzed below.

\subsection{Solution $S_{ab}$: $a>0$, $b<1$}
\label{sec:S_ab}

The final, more general case includes generic $a$ and $b$. In this case the resulting optimal eigenvalue density is given by
\begin{IEEEeqnarray}{rCl}
p(x)&=&\frac{\sqrt{(x-a)(b-x)}}{2\pi (1+\rho x)}\bigg(\frac{n_{0}(\rho +1)}{(1-x)\sqrt{(1-a)(1-b)}}\nonumber\\
&&+\frac{\beta -1}{x\sqrt{a b}}\bigg)
\end{IEEEeqnarray}
with the additional constraint
\begin{equation}\label{eq:p(b)constraint}
  \frac{n_0}{\sqrt{(1-a)(1-b)}}=\frac{\beta-1}{\sqrt{ab}}+\frac{k\rho}{\sqrt{(1+\rho a)(1+\rho b)}}
\end{equation}
obtained by demanding $p(a)=p(b)=0$. The parameters $a,b,k$ can be evaluated uniquely from the above equation in addition to the normalization constraint (\ref{eq:norm_condition})
\begin{equation}\label{eq:norm_condition_a>0b<1}
  n_0+\beta+1+k=\frac{\beta-1}{\sqrt{ab}}+\frac{k(1+\rho)}{\sqrt{(1+\rho a)(1+\rho b)}}
\end{equation}
and the rate constraint (\ref{eq:mutual_information_condition})
\begin{IEEEeqnarray}{rCl}\label{eq:r_eq}
r&=& r(k) \equiv \log{\Delta\rho}+
\frac{n_0}{2\sqrt{\barac\barbc}}\bigg[G\left(\baraz,\baraz\right)-G\left(\baraz,-\barac\right)\bigg] \\ \nonumber
&+&\frac{(\beta -1)}{2\sqrt{\bara \barb}}\bigg[G\left(\baraz,\bara\right)-G\left(\baraz,\baraz\right)\bigg]
\end{IEEEeqnarray}
where $\Delta=b-a$, $z=\frac{1}{\rho}$. For notational simplicity we also define $\bara=a/\Delta$, $\ac = 1-a$, $\barac=\ac/\Delta$, $\baraz=(a+z)/\Delta$ and $\barb=b/\Delta$, $\barbc=\bc/\Delta=(1-b)/\Delta$, $\barbz=(b+z)/\Delta$. The $G$ function can be seen in Appendix \ref{app:Gfn_def}. We may now integrate over $p(x)$ and obtain an expression for $\cE(r)$ as follows
\begin{IEEEeqnarray}{rCl}
\mathcal{E}(r)&=&\frac{k}{2}\left(r-\log\left(1+b\rho\right)\right)-\frac{\log\Delta}{2}\left(n_{0}+\beta+1 \right)-\frac{n}{2}\log{\bc}\nonumber\\
&-&\frac{n^2}{4\sqrt{\barac\barbc}}\left(G(\barbc,\barbc)-G(\barbc,-\barbz)\right)-\frac{(\beta -1)}{2}\log{b}\nonumber\\
&+&\frac{n(\beta -1)}{4\sqrt{\bara\barb}}\left(G(\barbc,-\barb)-G(\barbc,-\barbz)\right)\nonumber\\
&+&\frac{n(\beta -1)}{4\sqrt{\barac\barbc}}\left(G(\bara,-\barac)-G(\bara,\baraz)\right)\nonumber\\
&-&\frac{(\beta-1)^2}{4\sqrt{\bara\barb}} \left(G(\bara,\bara)-G(\bara,\baraz)\right) \nonumber \\
&-&\frac{n}{2\sqrt{\barac\barbc}} \left(G(0,\barbc)-G(0,-\barbz)\right) \nonumber \\
&+& \frac{\beta-1}{2\sqrt{\bara\barb}}\left(G(0,-\barb)-G(0,-\barbz)\right)
\end{IEEEeqnarray}

To make contact with the solutions of the previous sections, we observe that the conditions (\ref{eq:p(b)constraint}) and (\ref{eq:norm_condition_a>0b<1}) cannot be simultaneously be satisfied if $\beta=1$, $n>0$, and $k<k_{c3}$ (corresponding to $r<r_{c3}$) unless $a<0$. In this parameter region $S_{0b}$ applies. Also, for $\beta>1$, $n=0$, and $k>k_{c4}$ (and correspondingly $r>r_{c4}$), the above equations result to $b>1$, thereby invalidating the solution and necessitating the solution $S_{a1}$. In conclusion, we see that the above four solutions are mutually exclusive and cover all possible parameter values, thereby providing the unique solution to the exponent ${\cal E}(r)$ of the outage probability.

%
\begin{figure}
\centering{%
\includegraphics[scale=0.22]{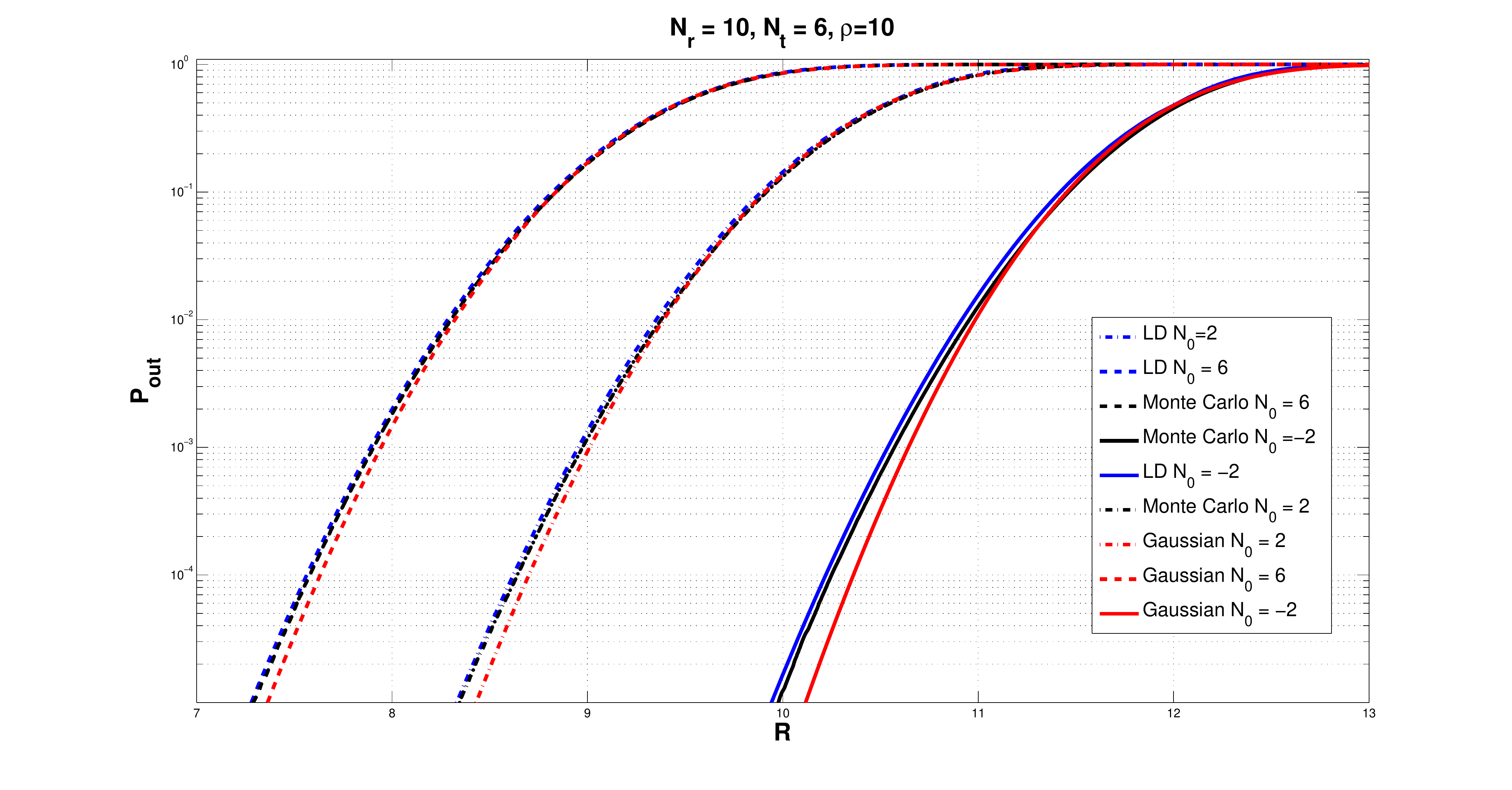}%
\caption{Outage probability curves for different values of $\textit{N}_0$. We observe that, generally,the Gaussian curves fail to follow the respective Monte Carlo, while the LD curves are closer to them.}
\label{fig:outage_lowsnr}
}
\end{figure}

\begin{figure}
\centering{%
\includegraphics[scale=0.22]{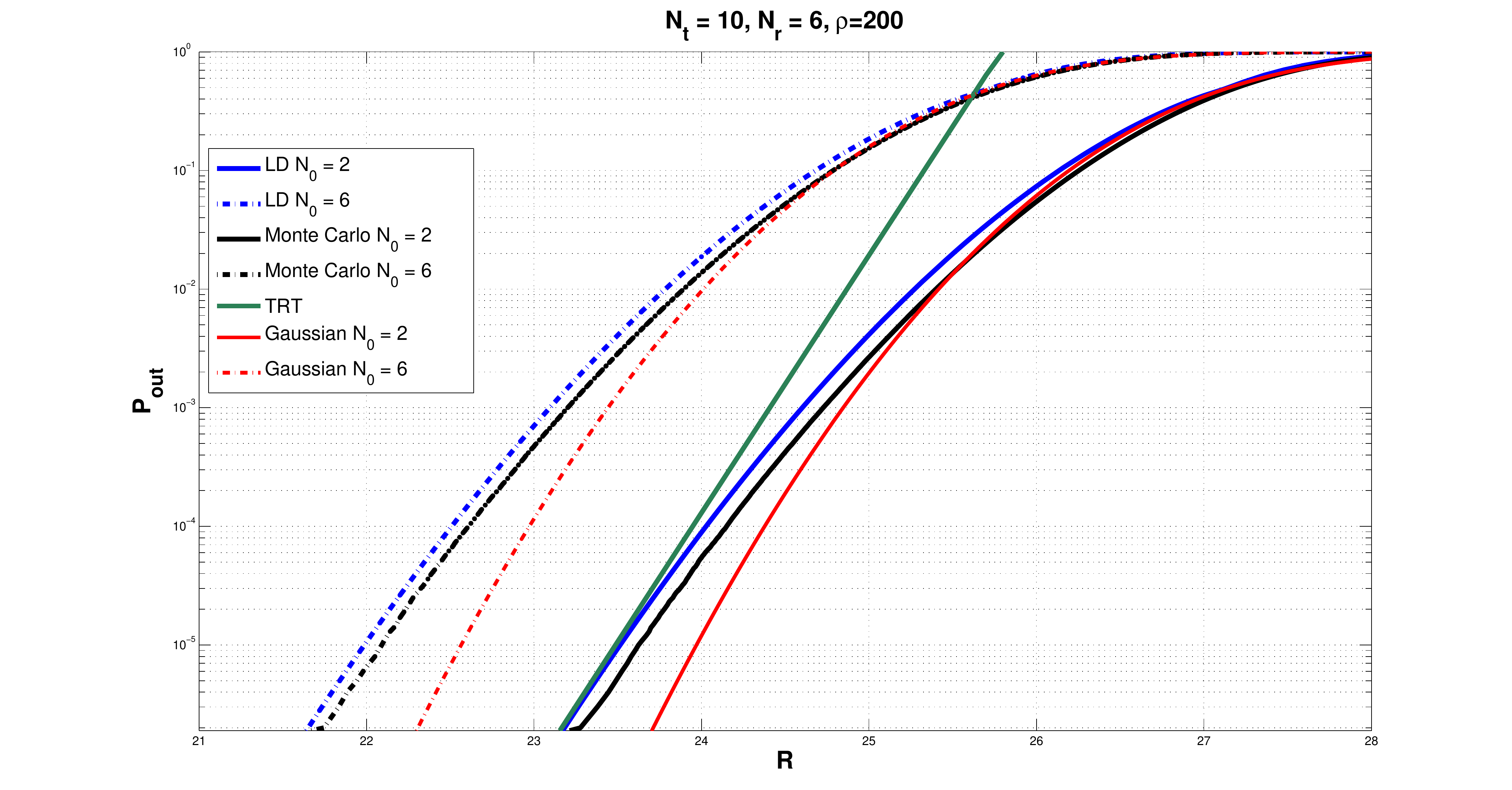}
\caption{For large SNR the Gaussian approximation does not provide such good results as the LD approach.}
\label{fig:outage_highsnr}
}
\end{figure}

\subsection{Probability Distributions $P(r)$  and $P_{out}(r)$}

In the previous sections we obtained the asymptotic behavior of the outage probability in the large $N_t$ limit. We found that the outage probability is approximately $P(I_N\leq r)\sim \exp[-N_t^2({\cal E}(r)-{\cal E}_0)]$ when $r<r_{erg}$ and we can similarly find for $r>r_{erg}$ that $P(I_N> r)\sim \exp[-N_t^2({\cal E}(r)-{\cal E}_0)]$. By differentiation we obtain to leading exponential order that the probability density follows the same law, i.e. $P(r)\sim \exp[-N_t^2({\cal E}(r)-{\cal E}_0)]$. To obtain the normalization constant for the density, we observe that the distribution close to its peak will be asymptotically Gaussian. This can be checked by calculating ${\cal E}(r)$ in the small $k$ limit and showing that it is quadratic in $r$. Hence the normalization of the distribution will be given for large $N_t$ by the variance of the distribution close to the peak. Therefore, we obtain
\begin{equation}\label{eq:PDF}
  P(r)\approx N_t\frac{e^{-N_t^2(\cE(r)-\cE_0)}}{\sqrt{2\pi v_{erg}}}
\end{equation}
where $v_{erg}$ is the variance at the peak of the distribution, and $r_{erg}$ is the solution of (\ref{eq:r_eq}) for $k=0$ corresponding to the ergodic rate. To obtain the value for $v_{erg}$ we observe that ${\cal E}'(r) = k(r)$, which is negative for $r<r_{erg}$ and positive for $r>r_{erg}$. Similarly, we can obtain the local variance by differentiating once again ${\cal E}''(r) = dk(r)/dr$. Setting $k=0$, it follows that
\begin{eqnarray}\label{eq:verg}
v_{erg} &=& \int_{a_0}^{b_0} dx \frac{dp(x,k)}{dk}\log(1+\rho x)  \nonumber \\
&=&\log{\frac{(\sqrt{1+\rho b_0}+\sqrt{1+\rho a_0})^2}{4\sqrt{1+\rho b_0}\sqrt{1+\rho a_0} }}
\end{eqnarray}
where $a_0$, $b_0$ are given in (\ref{eq:a0_b0_def}).

To obtain an expression for the outage probability that is continuous at $k=0$, we may integrate $P(r)$ above from $0$ to $r$ and noticing that due to the exponential dependence on $N_t$, only the region close to $r$ will be important. Thus for $r < r_{erg}$ the outage probability is
\begin{equation}
P_{out}(r)\approx\frac{e^{-N^{2}[\mathcal{E}_{1}(r)-\mathcal{E}_{0}-\frac{\mathcal{E}'_{1}(r)^{2}}{2\mathcal{E}''_{1}(r)}]}Q\bigg(\frac{N |\mathcal{E}'_{1}(r)|}{\sqrt{\mathcal{E}''_{1}(r)}}\bigg)}{\sqrt{\mathcal{E}_{1}''(r)\upsilon_{erg}}}
\label{pout}
\end{equation}
and for $r > r_{erg}$ it is
\begin{equation}
P_{out}(r)\approx 1-\frac{e^{-N^{2}[\mathcal{E}_{1}(r)-\mathcal{E}_{0}-\frac{\mathcal{E}'_{1}(r)^{2}}{2\mathcal{E}''_{1}(r)}]}Q\bigg(\frac{N |\mathcal{E}'_{1}(r)|}{\sqrt{\mathcal{E}''_{1}(r)}}\bigg)}{\sqrt{\mathcal{E}_{1}''(r)\upsilon_{erg}}}
\end{equation}
where $\mathcal{E}'_{1}(r)=k(r)$ and $\mathcal{E}''_{1}(r)=k'(r)$ are the first and second derivative of $\mathcal{E}_{1}(r)$ with respect to $\textit{r}$ and $Q(x)=\int_x^{\infty}dte^{-t^2/2}/\sqrt{2\pi}$. This approximation, while is essentially the same as (\ref{eq:P_N(r)}) when $N_t$ is large irrespective of $r$, but it is convenient, because it gives the crossover for fixed $N_t$ and $r\approx r_{erg}$.

\section{Numerical Simulations}
\label{sec:Simulations}

\begin{figure}[htp]
\centering{
\includegraphics[scale=0.22]{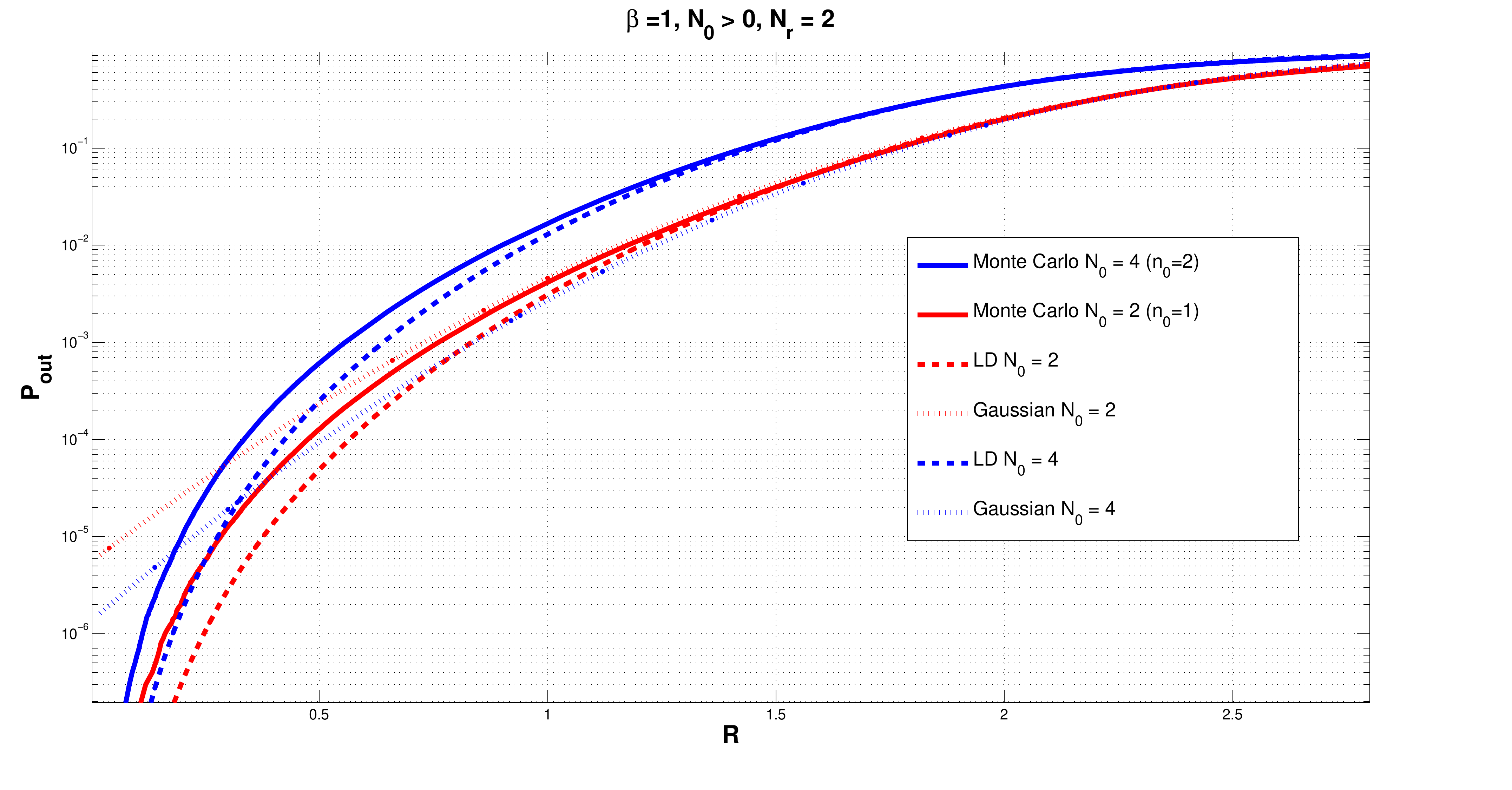}%
\caption{Simulation results for $\beta=1$}
\label{fig:beta1}
}
\end{figure}

To test the validity of the resulting equations above, we have performed a series of numerical simulations and have compared the $\textit{Large Deviation}$ (LD) approach to $\textit{Gaussian}$ approximation and $\textit{Monte Carlo}$ simulation. The Gaussian approximation consists of plotting $Q((R-N_t r_{erg})/v_{erg})$ versus $R$.
We plot indicative results for small and large $\rho$ (Figs. \ref{fig:outage_lowsnr} and \ref{fig:outage_highsnr}). It becomes clear that for large $\rho$ the Gaussian approximation does not perform well. Nevertheless the LD curves match Monte Carlo simulations even for small channel numbers. This difference in agreement holds also for $\beta=1$ (Fig. \ref{fig:beta1}).


\section{Conclusion}
\label{sec:Conclusion}

The purpose of this paper was to analyze the outage capacity for a particular model of the optical MIMO channel applicable to a multimode-multicore optical fiber system. The assumptions underlying the model assume strong forward scattering of light between the modes/cores, while the backscattering is weak. At the same time, we can model loss inside the fiber by varying a particular parameter of the model, namely $N_0$. We have provided two complementary approaches to provide analytic solutions for the outage capacity. In the first, we derived closed-form expressions for the outage probability. Despite its exactness, this approach becomes cumbersome to use beyond the size of a few channels. Therefore, we also implemented a large deviation approach first introduced in physics \cite{Dyson1962_DysonGas} to calculate the outage capacity for the optical MIMO channel in the limit of large channel numbers. Our method is especially applicable for the tails of the distribution, which is relevant for low outage requirements due to the absence of feedback and finite SNR. Our analytical results agree very well with numerical experiments. On the other hand the Gaussian approximation fails to follow the respective numerical and the deviation becomes greater as our system increases in size and complexity ($\beta$ and $n_0$) Additionally the method provides the distribution of eigenvalues constrained on the transmission rate and SNR. Although the channel assumptions taken here are somewhat idealized, this result gives an analytic metric to compare with other more complicated channel models. Clearly, more work is necessary, both from the channel sounding side, but also from the channel modeling side, so that the model will become more realistic.

\appendices

\section{Details for derivation of closed form solution}
\label{app:ExactSolution}

In this appendix we provide details of the derivation of the closed form expression for the outage presented in Section \ref{sec:ExactSolution}. We start with \eqref{eq:Pout_Fourier_xform1}. It is convenient to make the change of variables $1+\rho\lambda_k=y_k$ to get
\begin{align}
\nonumber 1-P_{out}(r) &={\cal A}\int_{-\infty}^\infty \frac{dp}{2\pi}\ \frac{e^{\mathrm{i}p N_tr}}{\epsilon-ip}\int_{[1,1+\rho]^{N_t}}d\by \prod_{n<m}|y_n-y_m|^2\\
&
\nonumber\times\prod_k \left[y_k^{-\mathrm{i}p}(y_k-1)^{|N_t-N_r|}\left((\rho+1)-y_k\right)^{N_0}\right]\
\\
&\nonumber {\cal A}= \frac{1}{Z_{N}\rho^{N_t^2+(|N_t-N_r|+N_0)N_t}}
\end{align}
where $d\by=dy_1\cdots dy_{N_t}$ and ${\cal A}$ a normalization constant. Now we invoke the Andr\'eief identity (see also Lemma 8 in \cite{Simon2006_UnitaryPaper}), which takes advantage of the fact that the products of the form $\prod_{n,m} (y_n-y_m)$ can be written as a Vandermonde determinant. Defining the function
\begin{equation}\label{eq:function_def1}
  g(x,p)=x^{-\mathrm{i}p}(x-1)^{|N_t-N_r|}\left((\rho+1)-x\right)^{N_0}
\end{equation}
we have
\begin{align}
\nonumber
& \int_{[1,1+\rho]^{N_t}}d\by \prod_{k=1}^{N_t} g(y_k,p) \det\left(y_i^{j-1}\right)^2 \\
& \nonumber %
= \int_{[1,1+\rho]^{N_t}}d\by \prod_{k=1}^{N_t} g(y_k,p) \left(\sum_{\ba(N_t)} (-1)^{|\ba|} \prod_{i=1}^{N_t} y_{i}^{a_i-1}\right)^2 \\
& \nonumber %
=  \sum_{\ba(N_t),{\bf b}(N_t)} (-1)^{|\ba|+|{\bf b}|} \int_{[1,1+\rho]^{N_t}}d\by \prod_{k=1}^{N_t} g(y_k,p) y_{k}^{a_k+b_k-2} \\
& \nonumber %
=  N_t! \det\left(H_{i+j}(p)\right)_{i,j=1,\ldots,N_t}
\label{phis}
\end{align}
In the second line we used the Leibnitz expansion of determinants\cite{Simon2006_UnitaryPaper}, where the sum is over all permutations $\ba$ with $(-1)^{|\ba|}$ being the sign of the permutation. In the final line we re-summed the integrated quantities to get a determinant of the Hankel matrix $\bH$ with elements $H_{i+j}$ given by
\begin{equation}
H_\ell(p)=\int_{1}^{1+\rho} dx\ x^{\ell-2-\mathrm{i}p}(x-1)^{|N_t-N_r|}((1+\rho)-x)^{N_0}\label{defint}
\end{equation}
resulting to the following expression for the outage probability
\begin{equation}
1-P_{out}(r)={\cal A}'\int_{-\infty}^\infty \frac{dp}{2\pi}\ \frac{e^{\mathrm{i}p N_t r}}{\epsilon-ip}\det\left(H_{i+j}(p)\right)_{i,j=1,\ldots,N_t}\label{phis}
\end{equation}
where ${\cal A}'=N_t!{\cal A}$.

The integral \eqref{defint} can be evaluated in the most elementary form exploiting the fact that both $|N_t-N_r|$ and $N_0$ are integers. Using the binomial theorem to expand the second and third powers, we get
\begin{align}
H_\ell(p)&=\sum_{k=0}^{|N_t-N_r|}\sum_{n=0}^{N_0}c_{k,n}
\frac{(1+\rho )^{\ell-1-\mathrm{i}p+k+N_0-n}-1}{\ell-1-\mathrm{i}p+k+N_0-n}
\label{defHs}
\end{align}
where
\begin{equation}\label{eq:c_kn}
c_{k,n}=\binom{|N_t-N_r|}{k}\binom{N_0}{n}(-1)^{|N_t-N_r|-k+N_0-n}(1+\rho)^{n}
\end{equation}
Let us now expand the determinant in \eqref{phis} using \eqref{defHs}. After rearranging the sums we get
\begin{equation}
1-P_{out}(r)=\sum_{\bk, \bn} C_{\bk,\bn} \sum_{\bsigma(N_t)}(-1)^{|\bsigma|} J\left(r; \{\mathbf{s}_{\bsigma}\}\right)
\end{equation}
where the sum over the integer components of the vector $\bk=[k_1,\cdots,k_{N_t}]$ is over the interval $[0, |N_t-N_r|]$, while for the vector $\bn=[n_1,\cdots,n_{N_t}]$ its components are summed over the interval $[0,N_0]$. Also, $C_{\bk,\bn}={\cal A}'\prod_i c_{k_i,n_i}$ and
\begin{equation}
J(r; \{\mathbf{s}_{\bsigma}\})=\int_{-\infty}^\infty \frac{dp}{2\pi}\ \frac{e^{\mathrm{i}p N_t r}}{\epsilon-ip} \prod_{j=1}^{N_t}\frac{(1+\rho )^{s_j-\mathrm{i}p}-1}
{s_j-\mathrm{i}p}\label{contour}
\end{equation}
where the components of the integer vector $\bs_{\bsigma}$ are $s_j=j+\sigma_j-1+k_j+N_0-n_j$. Expanding the numerator of the above equation, we obtain
\begin{align}
\nonumber J(r; \{\mathbf{s}\})&=\sum_{\ell=0}^{N_t}(-1)^{\ell+N_t} d_\ell(\bs) F(N_t r-\ell\log(1+\rho),\bs) \\
d_\ell(\bs) &= \be_\ell\left((1+\rho)^{s_1},\ldots,(1+\rho)^{s_{N_t}}\right) \\  %
F(z,\bs) &= \int_{-\infty}^\infty \frac{dp}{2\pi}\ \frac{e^{\mathrm{i}p z}}{(\epsilon-ip)\prod_{i=1}^{N_t}(s_i-\mathrm{i}p)}
\end{align}
where in the second line we have used the elementary symmetric polynomials $\be_\ell(x_1,x_2,\cdots,x_{N_t})$ of degree $\ell$.

As a result, in order to evaluate the outage probability in closed form we only need to evaluate the complex integral in $F(z,\bs)$. Since all poles of the integrand are in the lower half complex $p$-plane, if $z>0$ (hence $N_t r>\ell\log(1+\rho)$) then the integral vanishes \cite{Carrier_Krook_Pearson_Book_complex_analysis}. Hence only $\ell$-terms with $N_t r<\ell\log(1+\rho)$ survive. Having this in mind the integral can be evaluated by summing over the residues of the poles. As a result we obtain
\begin{eqnarray}
F(z,\bs)&=& \prod_{j=1}^{N_t} s_j^{-1} + \sum_{j=1}^{N_t}\frac{e^{s_jz}}{s_j\prod_{k\neq j}(s_k-s_j)} \\ \nonumber
  &=& \prod_{j=1}^{N_t} s_j^{-1} +  F_1(z,\bs)
\end{eqnarray}
Putting all above formulae together provides the final result expressed in \eqref{eq:exact_result_final1}.

Before concluding this section, it is worth discussing the value of the above equation when two or more integers $s_i$ are equal. To address this issue it will prove useful to express $F_1(z,\bs)$ as a ratio of determinants \cite{Kiessling2003_AnalyticalMIMOMMSE_CorrelatedFading}. Indeed we get
\begin{eqnarray}
F_1(z,\bs)= \frac{\det\left(f_i(s_j,z)\right)}{\prod_{n>m} (s_n-s_m)}
\end{eqnarray}
where the elements of the vector function ${\bf f}(x,z)$ is defined as follows
\begin{equation}
f_i(x,z) = \left\{\begin{array}{lr}
\frac{e^{xz}}{x} & i=1 \\
x^{i-1} & N_t\geq i>1
\end{array}\right.
\end{equation}
When one or more values of $s_j$ are identical, the ratio is ill-defined, because both numerator and denominator vanish. Although we could have dealt with the problem directly at the level of complex integration by considering double poles, it is more instructive to analyze this case as a limit of the $s$'s approaching each other. Following Lemma 1 in \cite{Moustakas2004_DUSTM} we can show that if $s_1$ has multiplicity $m$ then $F_1(z,\bs)$ can be expressed as
\begin{eqnarray}
F_1(z,\bs)= \frac{\det\bZ}{\prod_{a>b> m}(s_a-s_b)\prod_{j=m+1}^{N_t}(s_j-s_1)^m\prod_{q=1}^{m-1}q!}
\end{eqnarray}
where the matrix $\bZ$ can be expressed as
\begin{align}
\bZ= \left[\vec f(s_1,z);\right. &\vec f'(s_1,z);\ldots;\vec f^{(m-1)}(s_1,z);\\ \nonumber
&\left.\vec f(s_{m+1},z);\ldots;\vec f(s_{N_t},z)\right]
\end{align}
where the primes represent partial derivative with respect to the first argument.
We can similarly obtain expressions for the case when we have several multiplicities in $\bs$.

\section{Proof of Theorem \ref{thm:Coulomb_gas}}
\label{app:proof_LD}

In this appendix we will provide some details on the proof of the above theorem.

\subsubsection{Convexity}
The convexity of ${\cal E}[p]$ has been shown in \cite{BenArous1997_LDWignerLaw} over functions in ${\cal X}$, as also in \cite{Kazakopoulos2011_LivingAtTheEdge_LD_MIMO}.

\subsubsection{Uniqueness}
\label{app:E_0}

The uniqueness of the minimum of ${\cal E}[p]$ has been shown in \cite{BenArous1997_LDWignerLaw}\cite{HiaiPetz1998_LargeDeviationsWishartEigenvalues}.
The value of ${\cal E}_0$ can be obtained form the limit ${\cal E}_0 = -\lim_{N_t\rightarrow\infty}\log {\cal Z}_{N_t}/N_t^2$. However, the normalization factor ${\cal Z}_{N_t}$ can be evaluated explicitly using the Selberg integral \cite{Forrester2010_book} as follows:
\begin{eqnarray}\label{eq:ZN_def}
  {\cal Z}_{N_t} = \prod_{k=0}^{N-1} \frac{\Gamma(N(\beta-1)+1+k)\Gamma(Nn+1+k)\Gamma(k+2)}{\Gamma(N(\beta+n)+k+1)}
\end{eqnarray}
Using the Stirling approximation for the $\Gamma$-functions and approximating the sums with integrals, we get that
\begin{eqnarray}\label{eq:appE0}
  {\cal E}_0 &=& \frac{(\beta+n+1)^2}{2}\log(\beta+n+1)-\frac{(\beta+n)^2}{2}\log(\beta+n)\nonumber \\
  &-&\frac{\beta^2}{2}\log\beta + \frac{(\beta-1)^2}{2}\log(\beta-1) \nonumber \\
  &-&\frac{(1+n)^2}{2}\log(1+n) +\frac{n^2}{2}\log n
\end{eqnarray}

\subsubsection{Exponential Asymptote of $Prob(I_N<N_t r)$}

Let ${\cal X}_r$ be the set given by
\begin{eqnarray}\label{eq:Xr_def1}
\mathcal{X}_r = \left\{p\in {\cal X} \text{ and } \int_0^1\! p(x)\log(1+\rho x) \dd x \leq r \right\}
\end{eqnarray}
Given the linearity of the constraint, the above set is convex. Now, in \cite{HiaiPetz1998_LargeDeviationsWishartEigenvalues} it has been shown that $Prob(I_N\leq  r)$ obeys the large deviation principle with good rate function $I[p]={\cal E}[p]-{\cal E}_0$. Hence,
\begin{eqnarray}
{\cal E}_0-\inf_{p\in{\cal X}_r} {\cal E}[p]    &=& -\limsup_{N_t\rightarrow \infty} \frac{1}{N_t^2}\log P({\cal X}_r) \\ \nonumber
                                                &=& -\liminf_{N_t\rightarrow \infty} \frac{1}{N_t^2}\log P({\cal X}_r)
\end{eqnarray}

The analogous result can be obtained for Corollary \ref{cor:Coulomb_gas_complement} by noting that the complement of ${\cal X}_r$, namely ${\cal X}_r^c$ is also convex. Then the above result follows directly for $P(I_N>r)$.

\section{Uniqueness of solution of \eqref{eq:norm_condition_a0b<1} }
\label{app:uniqueness_norm_condition}

In this appendix we will show the uniqueness of solution of the normalization equation \eqref{eq:norm_condition_a0b<1}
\begin{eqnarray}
\frac{n_0}{\sqrt{1-b}}+k\sqrt{\frac{k}{z+b}}=2+n_0+k \nonumber
\end{eqnarray}

The left hand side of the above equation can, also, be identified as the in-parenthesis element of the eigenvalues density equation \eqref{eq:p(x)_a=0,b<1} for $x=b$.
We can set
\begin{eqnarray}
f(b) = \frac{n_0}{\sqrt{1-b}}+k\frac{\sqrt{z}}{\sqrt{z+b}}\nonumber
\end{eqnarray}
and taking the first derivative
\begin{eqnarray}
f'(b) = \frac{n_0}{(1-b)^{3/2}}-\frac{k\sqrt{z}}{(z+b)^{3/2}}\nonumber
\end{eqnarray}
\begin{itemize}
\item If $k<0$ it is $f'(b)>0$ and so, $f(b)$ is monotonous and \eqref{eq:norm_condition_a0b<1} has unique solution
\item If $k >0$ we also need the second derivative
\begin{eqnarray}
f''(b) = \frac{3}{2}\frac{n}{(1-b)^{5/2}}+\frac{3}{2}\frac{k\sqrt{z}}{(\sqrt{z}+b)^{5/2}}>0 \nonumber
\end{eqnarray}
The minimum value of $f(b)$ can be found for $b=0$ equal to $f(b)_{min}=n_0+k<2+n_0+k$, which is the right hand side of the \eqref{eq:norm_condition_a0b<1}, and the maximum value is for $b=1$, equal to$f(b)_{max} \rightarrow\infty$.  Finally, because $f'(b)=0$ has one real root, we can visualize that again \eqref{eq:norm_condition_a0b<1} has a unique solution.
\end{itemize}
The same procedure can be used to derive the respective solution uniqueness for the other cases.

\section{$G(x,y)$ and $I_3(x)$ Function}
\label{app:Gfn_def}

The function $G(x,y)$ for $x>0$ and $y>0$ or $y<-1$ is given by \cite{Kazakopoulos2011_LivingAtTheEdge_LD_MIMO}
\begin{IEEEeqnarray}{rCl}
&&G(x,y)=\frac{1}{\pi}\int_{0}^{1}\sqrt{t(1-t)}\frac{\log(t+x)}{t+y}d t\\ \nonumber
&=&-2\sgn(y)\sqrt{|y(1+y)|}\log \left[\frac{\sqrt{x|1+y|}+\sqrt{|y|(1+x)}}{\sqrt{|1+y|}+\sqrt{|y|}}\right]\nonumber\\
&+&(1+2 y)\log\left[\frac{\sqrt{1+x}+\sqrt{x}}{2}\right] -\frac{1}{2}\left(\sqrt{1+x}-\sqrt{x}\right)^{2}\nonumber
\end{IEEEeqnarray}
and $I_3(x)=-G(x,-1)$

\section*{Acknowledgment}
ALM would like to thank M. Feder for bringing this application of unitary matrices to his attention.


\begin{thebibliography}{10}
\providecommand{\url}[1]{#1}
\csname url@samestyle\endcsname
\providecommand{\newblock}{\relax}
\providecommand{\bibinfo}[2]{#2}
\providecommand{\BIBentrySTDinterwordspacing}{\spaceskip=0pt\relax}
\providecommand{\BIBentryALTinterwordstretchfactor}{4}
\providecommand{\BIBentryALTinterwordspacing}{\spaceskip=\fontdimen2\font plus
\BIBentryALTinterwordstretchfactor\fontdimen3\font minus
  \fontdimen4\font\relax}
\providecommand{\BIBforeignlanguage}[2]{{%
\expandafter\ifx\csname l@#1\endcsname\relax
\typeout{** WARNING: IEEEtran.bst: No hyphenation pattern has been}%
\typeout{** loaded for the language `#1'. Using the pattern for}%
\typeout{** the default language instead.}%
\else
\language=\csname l@#1\endcsname
\fi
#2}}
\providecommand{\BIBdecl}{\relax}
\BIBdecl
\renewcommand{\BIBentryALTinterwordstretchfactor}{4}

\bibitem{Scaling(Crunch)10}
R.~W. Tkach, ``Scaling optical communications for the next decade and beyond,''
  \emph{Bell Labs Technical Journal}, vol.~14, no.~4, pp. 3--9, 2010.

\bibitem{Winzer2011_OpticalMIMOCapacity}
P.~J. Winzer and G.~J. Foschini, ``{MIMO} capacities and outage probabilities
  in spatially multiplexed optical transport systems,'' \emph{Opt. Express},
  vol.~19, no.~17, pp. 16\,680--16\,696, Aug 2011.

\bibitem{Takenaga10CrosstalkMCF}
K.~Takenaga \emph{et~al.}, ``An investigation on crosstalk in multi-core fibers
  by introducing random fluctuation along longitudinal direction.'' \emph{IEICE
  Transactions}, vol. 94-B, no.~2, pp. 409--416, 2011.

\bibitem{Fini10CrosstalkBentMCF}
J.~M. Fini \emph{et~al.}, ``Statistics of crosstalk in bent multicore fibers,''
  \emph{Opt. Express}, vol.~18, no.~14, pp. 15\,122--15\,129, Jul 2010.

\bibitem{Hayashi11LowCrosstalkMCF}
T.~Hayashi \emph{et~al.}, ``Ultra-low-crosstalk multi-core fiber realizing
  space-division multiplexed ultra-long-haul transmission,'' in \emph{CLEO:
  Science and Innovations}.\hskip 1em plus 0.5em minus 0.4em\relax Optical
  Society of America, 2012, p. CTh4G.3.

\bibitem{Kato00TempChromDispers}
T.~Kato, Y.~Koyano, and M.~Nishimura, ``Temperature dependence of chromatic
  dispersion in various types of optical fiber,'' \emph{Optics Letters},
  vol.~25, 2000.

\bibitem{Zhu2010_7coreMCF}
B.~Zhu \emph{et~al.}, ``Seven-core multicore fiber transmissions for passive
  optical network,'' \emph{Opt. Express}, vol.~18, no.~11, pp.
  11\,117--11\,122, May 2010.

\bibitem{Morioka2012_CommMag_Enhancing_OCommsFibers}
T.~Morioka \emph{et~al.}, ``Enhancing optical communications with brand new
  fibers,'' \emph{Communications Magazine, IEEE}, vol.~50, no.~2, pp. s31
  --s42, february 2012.

\bibitem{TarighatFundamentals07}
A.~Tarighat \emph{et~al.}, ``Fundamentals and challenges of optical
  multiple-input multiple-output multimode fiber links [topics in optical
  communications],'' \emph{Communications Magazine, IEEE}, vol.~45, no.~5, pp.
  57 --63, may 2007.

\bibitem{Hsu06capacityenhancement}
R.~C.~J. Hsu \emph{et~al.}, ``Capacity enhancement in coherent optical mimo
  (comimo) multimode fiber links,'' \emph{IEEE Comm. Letters}, vol.~10, pp.
  195--197, 2006.

\bibitem{Jackson_EM_book}
J.~D. Jackson, \emph{Classical Electrodynamics}, 3rd~ed.\hskip 1em plus 0.5em
  minus 0.4em\relax New York: J. Wiley \& Son, Inc., 1998.

\bibitem{Simon2006_UnitaryPaper}
S.~H. Simon and A.~L. Moustakas, ``Crossover from conserving to lossy in
  circular random matrix ensembles,'' \emph{Physical Review Letters}, vol.~96,
  no.~13, p. 136805, 2006.

\bibitem{Forrester2010_book}
P.~J. Forrester, \emph{Log-gases and Random Matrices}.\hskip 1em plus 0.5em
  minus 0.4em\relax Princeton, NJ: Princeton Univ. Press, 2010.

\bibitem{Shah05coherentoptical}
A.~R. Shah \emph{et~al.}, ``Coherent optical {MIMO} ({COMIMO}),'' \emph{IEEE J.
  Lightwave Tech}, vol.~23, pp. 2410--2419, 2005.

\bibitem{Dar2012_JacobiMIMOChannel}
R.~Dar, M.~Feder, and M.~Shtaif, ``The {J}acobi {MIMO} channel,'' \emph{CoRR},
  vol. abs/1202.0305, 2012.

\bibitem{Dyson1962_DysonGas}
F.~Dyson, ``Statistical theory of the energy levels of complex systems. {I},''
  \emph{J. Math. Phys.}, vol.~3, p. 140, 1962.

\bibitem{Majumdar2006_LesHouches}
S.~N. Majumdar, \emph{Random Matrices, the {U}lam Problem, Directed Polymers \&
  Growth Models, and Sequence Matching}, ser. Les Houches, M.~M{\'e}zard and
  J.~P. Bouchaud, Eds.\hskip 1em plus 0.5em minus 0.4em\relax Elsevier, July
  2006, vol. Complex Systems.

\bibitem{Vivo2007_LargeDeviationsMaxEigvalueWishart}
P.~Vivo, S.~N. Majumdar, and O.~Bohigas, ``Large deviations of the maximum
  eigenvalue in {W}ishart random matrices,'' \emph{Journal of Physics A:
  Mathematical and Theoretical}, vol.~40, no.~16, pp. 4317--4337, 2007.

\bibitem{Dean2008_ExtremeValueStatisticsEigsGaussianRMT}
D.~S. Dean and S.~N. Majumdar, ``Extreme value statistics of eigenvalues of
  {G}aussian random matrices,'' \emph{Phys. Rev E}, vol.~77, p. 041108, 2008.

\bibitem{Vivo2008_DistributionsConductanceShotNoise}
P.~Vivo, S.~N. Majumdar, and O.~Bohigas, ``Distributions of conductance and
  shot noise and associated phase transitions,'' \emph{Phys. Rev. Lett.}, vol.
  101, p. 216809, 2008.

\bibitem{Kazakopoulos2011_LivingAtTheEdge_LD_MIMO}
P.~Kazakopoulos \emph{et~al.}, ``Living at the edge: A large deviations
  approach to the outage mimo capacity,'' \emph{Information Theory, IEEE
  Transactions on}, vol.~57, no.~4, pp. 1984 --2007, april 2011.

\bibitem{Chen2012_CoulombMIMO}
Y.~Chen and M.~R. McKay, ``Coulomb fluid, {P}ainleve transcendents and the
  information theory of {MIMO} systems,'' \emph{IEEE Transactions on
  Information Theory}, vol.~58, pp. 4594--4634, Jul. 2012.

\bibitem{Beenakker1997_MesoscopicReview}
C.~W.~J. Beenakker, ``Random-matrix theory of quantum transport,'' \emph{Rev.
  Mod. Phys.}, vol.~69, pp. 731--808, 1997.

\bibitem{Martin1992_Landauer}
T.~Martin and R.~Landauer, ``Wave-packet approach to noise in multichannel
  mesoscopic systems,'' \emph{Phys. Rev. {\bf B}}, vol.~45, no.~4, pp.
  1742--1755, 1992.

\bibitem{Gysel1990_StatisticalPropertiesRayleighScettering}
P.~Gysel and R.~K. Staubli, ``Statistical properties of rayleigh backscattering
  in single-mode fibers,'' \emph{IEEE Journal of Lightwave Technology}, vol.~8,
  1990.

\bibitem{Gifford2012_OpticalBackscatterReflectometry}
D.~Gifford, ``Rayleigh backscatter reflectometry boosts fiber
  characterization,'' \emph{Laser Focus World}, June 2012.

\bibitem{Martinez2010_BackscateredOpticalPowerBidirectionPON}
J.~Martínez \emph{et~al.}, ``Analysis of the influence of backscattered
  optical power over bidirectional {PON} links,'' \emph{Optics Communications},
  vol. 283, 2010.

\bibitem{Ohasi2012_BackscatteredPower_CrosstalkMultiCoreFibers}
M.~Ohashi \emph{et~al.}, ``Simple backscattered power technique for measuring
  crosstalk of multi-core fibers,'' \emph{17th Opto-Electronics and
  Communications Conference, Busan, Korea}, 2012.

\bibitem{Ryf2013_CombinedSDMWDM_FMF}
R.~Ryf \emph{et~al.}, ``Combined {SDM} and {WDM} transmission over 700-km
  few-mode fiber,'' \emph{OFC/NFOEC Technical Digest, OSA}, 2013.

\bibitem{Foschini1998_BLAST1}
G.~J. Foschini and M.~J. Gans, ``On limits of wireless communications in a
  fading environment when using multiple antennas,'' \emph{Wireless Personal
  Communications}, vol.~6, pp. 311--335, 1998.

\bibitem{Telatar1995_BLAST1}
I.~E. Telatar, ``Capacity of multi-antenna {G}aussian channels,''
  \emph{European Transactions on Telecommunications and Related Technologies},
  vol.~10, no.~6, pp. 585--596, Nov. 1999.

\bibitem{Wang2002_OutageMutualInfoOfSTMIMOChannels}
Z.~Wang and G.~B. Giannakis, ``Outage mutual information of space-time {MIMO}
  channels,'' \emph{{IEEE} Trans. Inform. Theory}, vol.~50, no.~4, pp.
  657--662, Apr. 2004.

\bibitem{Macdonald1995_SymmetricFunctions_book}
I.~G. Macdonald, \emph{Symmetric Functions and Hall Polynomials}, 2nd~ed.\hskip
  1em plus 0.5em minus 0.4em\relax Oxford: Clarendon Press, 1995.

\bibitem{Mehta_book}
M.~L. Mehta, \emph{Random Matrices}, 2nd~ed.\hskip 1em plus 0.5em minus
  0.4em\relax San Diego, CA: Academic Press, 1991.

\bibitem{BenArous1997_LDWignerLaw}
G.~{Ben Arous} and A.~Guionnet, ``Large deviations for the {W}igner's law and
  {V}oiculescu's non-commutative entropy,'' \emph{Prob. Theory Relat. Fields},
  vol. 108, pp. 517--542, 1997.

\bibitem{HiaiPetz1998_LargeDeviationsWishartEigenvalues}
F.~Hiai and D.~Petz, ``Eigenvalue density of the wishart matrix and large
  deviations,'' \emph{Infinite Dimensional Anal. Quantum Prob.}, vol.~1, pp.
  633--646, 1998.

\bibitem{Boyd_book}
S.~Boyd and L.~Vandenberghe, \emph{Convex Optimization}.\hskip 1em plus 0.5em
  minus 0.4em\relax Cambridge Univ. Press, 2004.

\bibitem{Tricomi_book_IntegralEquations}
F.~G. Tricomi, \emph{Integral Equations}, ser. Pure Appl. Math V.\hskip 1em
  plus 0.5em minus 0.4em\relax London: Interscience, 1957.

\bibitem{Debbah2003_UnitaryAsymptoticallyFreeMatrices}
M.~Debbah \emph{et~al.}, ``{MMSE} analysis of certain large isometric random
  precoded systems,'' \emph{{IEEE} Trans. Inform. Theory}, vol.~49, no.~5, p.
  1293, May 2003.

\bibitem{Carrier_Krook_Pearson_Book_complex_analysis}
G.~F. Carrier, M.~Krook, and C.~E. Pearson, \emph{Functions of a Complex
  Variable}.\hskip 1em plus 0.5em minus 0.4em\relax New York: McGraw-Hill,
  1966.

\bibitem{Kiessling2003_AnalyticalMIMOMMSE_CorrelatedFading}
M.~Kiessling and J.~Speidel, ``Analytical performance of {MIMO} {MMSE}
  receivers in correlated {R}ayleigh fading environments,'' in \emph{Vehicular
  Technology Conference, 2003. VTC 2003-Fall. 2003 IEEE 58th}, vol.~3.\hskip
  1em plus 0.5em minus 0.4em\relax IEEE, 2003, pp. 1738--1742.

\bibitem{Moustakas2004_DUSTM}
A.~L. Moustakas, S.~H. Simon, and T.~L. Marzetta, ``Capacity of differential
  versus non-differential unitary space-time modulation for {MIMO} channels,''
  \emph{{IEEE} Trans. Inform. Theory}, vol.~52, no.~8, pp. 3622--3634, Aug.
  2006.

\end{thebibliography}
\end{document}